\documentclass[aps,prd,nofootinbib]{revtex4}

\begin{document}
\title{Covariant description of parametrized nonrelativistic Hamiltonian
systems}
\date{\today}

\author{Mauricio Mondrag\'on}
\email{mo@fis.cinvestav.mx}
\affiliation{Departamento de F\'{\i}sica, Centro
de Investigaci\'on y de Estudios Avanzados del I.P.N., Av. I.P.N. No. 2508,
07000 Ciudad de M\'exico, M\'exico.}

\author{Merced Montesinos\footnote{Associate Member of the Abdus Salam
International Centre for Theoretical Physics, Trieste, Italy.}}
\email{merced@fis.cinvestav.mx}
\affiliation{Departamento de F\'{\i}sica,
Centro de Investigaci\'on y de Estudios Avanzados del I.P.N., Av. I.P.N. No.
2508, 07000 Ciudad de M\'exico, M\'exico.}

\begin{abstract}
The various phase spaces involved in the dynamics of parametrized
nonrelativistic Hamiltonian systems are displayed by using Crnkovic and
Witten's covariant canonical formalism. It is also pointed out that in Dirac's
canonical formalism there exists a freedom in the choice of the symplectic
structure on the extended phase space and in the choice of the equations that
define the constraint surface with the only restriction that these two choices
combine in such a way that any pair (of these two choices) generates the same
gauge transformation. The consequence of this freedom on the algebra of
observables is also discussed.
\end{abstract}


\maketitle

\section{Introduction}
There is currently a growing interest in the study of the fundamentals of both
classical and quantum mechanics motivated, in part, by several theoretical
approaches that try to build a quantum theory of gravity [see, for instance,
Ref. \cite{RovJMP}]. The various conceptual issues found in the construction
of generally covariant quantum theories frequently make people to go back to
the fundamentals of both classical and quantum mechanics to try to remove what
is non-essential and get the generic aspects of them which could be
implemented later on in realistic theories [see, for instance, Refs.
\cite{Rovelli1999,Smolin,RovPO,Mike,Marolf} and references therein].

In this paper we focus in the covariant description of Hamiltonian mechanics.
The geometrical structure underlying parametrized nonrelativistic Hamiltonian
systems is obtained by using the approach of Ref. \cite{Witten}, from which
the extended phase space $(\Gamma_{ext}, \Omega_{ext})$ and the presymplectic
phase space $(\Sigma, \Omega_{\Sigma})$ involved are obtained [see also Ref.
\cite{MO} for more details]. Once this is done, the definition of physical
observables is implemented and this fact allows us to reach the physical phase
space $(\Gamma_{phys}, \Omega_{phys})$ for the system. This is displayed in
SubSecs. \ref{Hamilton} and \ref{Geometry}. In spite of working with the
covariant canonical formalism, the usual symplectic structure is used. The
implications of choosing alternative symplectic structures in Dirac's
formalism are analyzed in Secs. III, IV, V, and VI where it is shown that
there are many ways of choosing the symplectic structure on the extended phase
space if the equation that defines the constraint surface is, in the generic
case,\footnote{In some cases, the equation that defines the constraint surface
is not modified [see Sec. III]. } accordingly modified in such a way that the
gauge transformation is not altered. Due to the fact that the gauge
transformation is not modified the gauge-invariant functions are also not
modified, however, the `algebra of observables' is, in the generic case,
modified because it depends on the particular symplectic structure chosen.
Section VII contains a generalization of these results to generally covariant
systems with first class constraints only. Our conclusions are collected in
Sec. VIII.

\section{The geometry and the physics of parametrized nonrelativistic
Hamiltonian systems}
Let us begin by considering the Hamiltonian formulation
associated with a system with a finite number of degrees of freedom obtained
from the action principle
\begin{eqnarray}
S[q^j,p_j] & = & \int^{t_2}_{t_1} \left [ \frac{d q^j}{dt}p_j - H (q^j ,
p_j,t) \right ] dt \, , \quad j=1,..., n,  \label{NonPar}
\end{eqnarray}
subject to the standard boundary conditions
\begin{eqnarray}
q^j ( t_{\alpha}) & = & q^j_{\alpha} \, , \quad \alpha=1,2,
\end{eqnarray}
where $q^j_{\alpha}$ are prescribed numbers. It is assumed that the dynamical
system is well-defined, i.e., that there exists a solution to the dynamical
problem that matches the boundary conditions. By definition, the coordinates
$q^i$ locally label the points of the {\it configuration space} ${\cal C}$ for
the system; the cotangent bundle of ${\cal C}$, $\Gamma=T^{\ast} {\cal C}$, is
the {\it phase space} whose points are locally labelled by the coordinates
$x^a$, $x^i=q^i$ and $x^{i+n}=p_i$, $a=1,...,2n$. The equations of motion for
the system have the canonical form
\begin{eqnarray}
{\dot x}^a & = & \omega^{ab} \frac{\partial H}{\partial x^b}\, , \label{Hform}
\end{eqnarray}
where $H$ is the Hamiltonian, $\omega^{ab}$ are the components of the inverse
of the symplectic matrix
\begin{eqnarray}
\{ x^a , x^b \} & := & \omega^{ab} \, , \label{Fbras} \quad (\omega^{ab} )  =
\left (
\begin{array}{rr}
0 & I \\
- I & 0
\end{array}
\right ) \, ,
\end{eqnarray}
with $I$ the $n\times n$ identity matrix and $0$ the $n\times n$ zero matrix.

The symplectic structure
\begin{eqnarray}
\omega & := & \frac12 \omega_{ab} d x^a \wedge d x^b \, , \label{SymEst}
\end{eqnarray}
induces a Poisson structure on $\Gamma=T^{\ast} {\cal C}$ defined by
\begin{eqnarray}
\{ f (x,t) , g(x,t) \} := \frac{\partial f}{\partial x^a} \omega^{ab}
\frac{\partial g}{\partial x^b} \, , \label{PoissDeff}
\end{eqnarray}
where, as usual, the coordinate $t$ is treated as a parameter \cite{Olver}.

{\it Parameterizing the system}. If the time variable is considered as a
canonical variable, the action for this Hamiltonian system becomes
\begin{eqnarray}
S[q^j,p_j,t,p_t, \lambda] & = & \int^{\tau_2}_{\tau_1} \left [{\dot q}^j p_j +
{\dot t} p_t - \lambda \gamma  \right ]d\tau \, ,
\nonumber\\
\gamma & := & p_t + H(q^j,p_j,t) \, , \label{HamActPar}
\end{eqnarray}
subject to the standard boundary conditions
\begin{eqnarray}
q^j (\tau_{\alpha}) & = & q^j_{\alpha} \, ,  \quad t(\tau_{\alpha}) =
t_{\alpha} \, , \quad \alpha=1,2,
\end{eqnarray}
where $p_t$ is the canonical variable conjugate to $t$, $\lambda$ is the
Lagrange multiplier associated with the constraint $\gamma=0$ that comes from
the definition of $p_t$, and the dot means total derivative with respect to
the parameter $\tau$ \footnote{It is frequently asserted that the parameter
$\tau$ has no `physical meaning.' However, this assertion is not completely
true. For instance, in the case of the relativistic free particle the
parameter $\tau$ might be chosen to be the proper time which, of course, has a
physical content; it is the reading of a clock moving together with the
relativistic particle, i.e., the proper time can be measured by a device.}.
The new configuration space for the system is the {\it extended configuration
space} ${\cal C}_{ext} = {\cal C}\times R$ whose points are locally labelled
by $(q^i,t)$ where $R$ stands for the $t$ coordinate. Its corresponding phase
space will be analyzed below.

\subsection{Hamilton's principle}\label{Hamilton}
Following the conventions used in Ref. \cite{Greiner}, let ${\tilde\delta}$
denote the arbitrary variation of coordinates $(q^j,t)$, momenta $(p_j,p_t)$,
and Lagrange multiplier $\lambda$ at $\tau$ fixed
\begin{eqnarray}
{\tilde\delta} x^{\mu} (\tau) & := & {x'}^{\mu} (\tau) - x^{\mu} (\tau)
\, , \nonumber\\
{\tilde\delta} \lambda (\tau) & := & {\lambda'}(\tau) -\lambda (\tau)\, ,
\label{totvar}
\end{eqnarray}
where $(x^{\mu})=(q^j , t, p_j, p_t)$, $\mu=1,...,2(n+1)$. The object
${\tilde\delta}$ is, but not always, called the {\it total variation}, other
authors call it a virtual variation.

The variation of the action (\ref{HamActPar}) under arbitrary transformations
(\ref{totvar}) keeping at most first order terms in ${\tilde\delta}$ is
\begin{eqnarray}
{\tilde\delta} S & := & S[{q'}^j,{p'}_j,{t'},{p'}_t,{\lambda'}]-S[q^j, p_j,
t,p_t,\lambda] \nonumber\\
& = & \int^{\tau_2}_{\tau_1} \left [ - \left ({\dot p}_t + \lambda
\frac{\partial \gamma }{\partial t} \right ){\tilde\delta} t + \left ({\dot
t}-\lambda \frac{\partial \gamma}{\partial p_t } \right ){\tilde\delta} p_t -
\left ( {\dot p}_j + \lambda \frac{\partial \gamma }{\partial q^j} \right )
{\tilde\delta} q^j \right. \nonumber\\
& & + \left. \left ( {\dot q}^j - \lambda \frac{\partial \gamma }{\partial
p_j} \right ){\tilde\delta} p_j - \gamma {\tilde\delta}\lambda +
\frac{d}{d\tau} \left ( p_t {\tilde\delta} t + p_j {\tilde\delta} q^j \right )
\right ] d\tau \, . \label{ActVar}
\end{eqnarray}
To get the equations of motion for the system, Hamilton's principle will be
used. According to it, the evolution of the system from $\tau_1$ to $\tau_2$,
keeping the end points fixed [$ {\tilde\delta} q^j(\tau_{\alpha})=0=
{\tilde\delta} t (\tau_{\alpha})$], is along the path such that the total
variation of the action vanishes, ${\tilde \delta} S=0$. On the other hand,
the variations ${\tilde\delta} t$, ${\tilde\delta} p_t$, ${\tilde\delta} q^j$,
${\tilde \delta} p_j$, and ${\tilde\delta} \lambda $ are arbitrary in the
`bulk' $(\tau_1, \tau_2)$. This, together with ${\tilde\delta}S=0$ and the
fact that ${\tilde\delta} q^j(\tau_{\alpha})=0= {\tilde\delta} t
(\tau_{\alpha})$ imply that the coefficients of the variations in the
integrand must vanish
\begin{eqnarray}
{\tilde\delta} q^j & : & {\dot p}_j = -
\lambda \frac{\partial \gamma}{\partial q^j}\, , \nonumber\\
{\tilde \delta} t & : & {\dot p}_t = -
\lambda \frac{\partial \gamma}{\partial t} \, , \nonumber\\
{\tilde \delta} p_j & : &  {\dot q}^j =
\lambda \frac{\partial \gamma}{\partial p_j } \, , \nonumber\\
{\tilde \delta } p_t & : & \,\,\, {\dot t} =
\lambda \frac{\partial \gamma}{\partial p_t} \, , \nonumber\\
{\tilde \delta} \lambda & : & \,\,\, \gamma =0 \, , \label{ParEqs}
\end{eqnarray}
which are the equations of motion for the parametrized Hamiltonian system
(\ref{HamActPar}).

Note that the Lagrange multiplier $\lambda$ {\it cannot} be determined from
the evolution of the constraint. To see this, the evolution with respect to
$\tau$ of $\gamma$ is computed. If $f=f(x^{\mu},\tau)$ then
\begin{eqnarray}
\frac{d}{d\tau} f(x^{\mu},\tau) & = & \frac{\partial f}{\partial q^j }{\dot
q}^j + \frac{\partial f}{\partial t} {\dot t}  + \frac{\partial f}{\partial
p_j} {\dot p}_j  + \frac{\partial f}{\partial p_t}{\dot p}_t + \frac{\partial
f}{\partial \tau}\, , \nonumber\\
& = & \lambda \{f, \gamma \}_{ext} + \frac{\partial f}{\partial \tau}\, ,
\end{eqnarray}
where the Poisson bracket $\{ \cdot , \cdot \}_{ext}$ is defined by
\begin{eqnarray}
\{ f , g \}_{ext} & := &  \frac{\partial f}{\partial q^i} \frac{\partial
g}{\partial p_i} + \frac{\partial f }{\partial t} \frac{\partial g}{\partial
p_t} - \left
(  f \longleftrightarrow g \right ) \, , \label{Poisson}\\
& \equiv & \frac{\partial f}{\partial x^{\mu}} \Omega^{\mu\nu}_{ext}
\frac{\partial g}{\partial x^{\nu}} \, ,
\end{eqnarray}
where $\Omega^{\mu\nu}_{ext}$ are the components of the inverse of the
symplectic matrix
\begin{eqnarray}
\{ x^{\mu}, x^{\nu} \}_{ext} & := & \Omega^{\mu\nu}_{ext} \, , \quad \left (
\Omega^{\mu\nu}_{ext} \right ) =  \left (
\begin{array}{rr}
0 & I \\
- I & 0
\end{array}
\right ) \, ,
\end{eqnarray}
where $I$ is the $(n+1)\times(n+1)$ identity matrix and $0$ the
$(n+1)\times(n+1)$ zero matrix.

In particular, if $f=\gamma$ then $\{\gamma , \gamma \}_{ext}$=0 and
$\frac{\partial \gamma }{\partial \tau}=0$ and so $\frac{d \gamma}{d\tau}=0$.
One could alternatively say that no more constraints arise and that $\gamma$
is the only one constraint in the formalism. In Dirac's terminology $\gamma$
is first class \cite{Dirac}.

\subsection{Extended, presymplectic, and physical phase spaces}\label{Geometry}
Let us denote the integrand of Eq. (\ref{HamActPar}) as
\begin{eqnarray}
L := {\dot q}^j p_j + {\dot t} p_t - \lambda \gamma \, .
\end{eqnarray}
Following Ref. \cite{Witten} and using Eq. (\ref{ActVar}) it follows that the
total variation of $L$ is
\begin{eqnarray}
{\tilde\delta L} & = & - \left ( {\dot p}_t + \lambda \frac{\partial \gamma
}{\partial t} \right ) {\tilde\delta} t + \left ({\dot t}-\lambda
\frac{\partial \gamma}{\partial p_t}\right ){\tilde\delta} p_t - \left ( {\dot
p}_j + \lambda \frac{\partial \gamma }{\partial q^j} \right ) {\tilde\delta}
q^j \nonumber\\
& & + \left ( {\dot q}^j - \lambda \frac{\partial \gamma }{\partial p_j}
\right ){\tilde\delta} p_j - \gamma {\tilde\delta}\lambda + \frac{d}{d\tau }
\left ( p_t {\tilde\delta} t + p_j {\tilde\delta} q^j \right )\, .
\end{eqnarray}
Computing again ${\tilde\delta}$ and using ${\tilde\delta}^2 L=0$
\begin{eqnarray}
\frac{d \Omega_{ext}}{d \tau} & = &  {\tilde\delta} \left ( {\dot p}_t +
\lambda \frac{\partial \gamma }{\partial t} \right ) \wedge {\tilde\delta} t -
{\tilde\delta} \left ({\dot t}-\lambda \frac{\partial \gamma}{\partial p_t}
\right ) \wedge {\tilde\delta} p_t + {\tilde\delta} \left ( {\dot p}_j +
\lambda \frac{\partial \gamma }{\partial q^j} \right ) \wedge {\tilde\delta}
q^j  \nonumber\\
& & + {\tilde\delta} \gamma \wedge {\tilde\delta} \lambda - {\tilde\delta}
\left ( {\dot q}^j - \lambda \frac{\partial \gamma }{\partial p_j} \right )
\wedge {\tilde\delta} p_j\, , \label{ContEq}
\end{eqnarray}
where
\begin{eqnarray}
\Omega_{ext} & := &  {\tilde\delta} p_i \wedge {\tilde\delta} q^i +
{\tilde\delta} p_t \wedge {\tilde\delta} t \, ,
 \label{symplectic}
\end{eqnarray}
[{\it cf.} Refs. \cite{RovPO,RovNote,RovNote3}].

{\it Definition}. The {\it extended phase space} for the system is the couple
$(\Gamma_{ext}, \Omega_{ext})$, where $\Gamma_{ext}$ is equal to the cotangent
bundle of the extended configuration space ${\cal C}_{ext}$,
$\Gamma_{ext}=T^{\ast} {\cal C}_{ext}= T^{\ast} {\cal C}\times R^2$. The
points of $\Gamma_{ext}$ can be labelled by $(x^{\mu})=(q^j, t, p_j, p_t)$.
The symplectic structure on $\Gamma_{ext}$ is given in Eq. (\ref{symplectic}),
which is the one already defined in Eq. (\ref{Poisson}). Note that
$\Omega_{ext}$ is closed and non-degenerate, i.e., it is a symplectic
structure on $\Gamma_{ext}$.

Note also that `on-shell', i.e., if the equations of motion (\ref{ParEqs})
hold, then their total variation vanishes too
\begin{eqnarray}
{\tilde\delta} \left ( {\dot p}_t + \lambda \frac{\partial \gamma }{\partial
t} \right ) & =  & 0 \, , \nonumber\\
{\tilde\delta} \left ({\dot t}-\lambda \frac{\partial \gamma}{\partial p_t}
\right ) & = & 0 \, , \nonumber\\
{\tilde\delta} \left ( {\dot p}_j + \lambda \frac{\partial \gamma }{\partial
q^j} \right ) & = & 0 \, ,\nonumber\\
{\tilde\delta } \left ( {\dot q}^j - \lambda \frac{\partial \gamma}{\partial
p_j} \right ) & = & 0 \, , \nonumber\\
{\tilde\delta} \gamma & = & 0 \, . \label{ContEqII}
\end{eqnarray}
Thus, from Eqs. (\ref{ContEq}), (\ref{symplectic}) and (\ref{ContEqII})
\begin{eqnarray}
\frac{d \Omega_{\Sigma}}{d\tau} & = & 0 \, , \label{Cons}
\end{eqnarray}
where
\begin{eqnarray}
\Omega_{\Sigma} & := & i^{\ast} \Omega_{ext} = {\tilde\delta} p_i \wedge
{\tilde\delta} q^i - {\tilde\delta} H(q,p,t) \wedge {\tilde\delta} t \, ,
\label{degene}
\end{eqnarray}
is the pull-back of $\Omega_{ext}$ to the constraint surface $\Sigma$ defined
by $\gamma=0$ through the inclusion map $i:\Sigma\longrightarrow \Gamma_{ext}$
[{\it cf}. Refs. \cite{RovPO,RovNote,RovNote3}]. To be precise, $\Sigma := \{
(q^i,t,p_i,p_t) \in \Gamma_{ext} \mid p_t=-H(q,p,t) \}$. It is clear that
$\Sigma$ is a $(2n+1)$-dimensional submanifold of $\Gamma_{ext}$ and that
$(y^A)=(q^i,p_i,t)$ can be taken as independent coordinates for labelling the
points on $\Sigma$. The inclusion map $i$ from $\Sigma$ to $\Gamma_{ext}$ is
defined by
\begin{eqnarray}
i & : & \Sigma \longrightarrow \Gamma_{ext}\, , \nonumber\\
 & & (q,p,t) \longmapsto (q,p,t,p_t=-H(q,p,t)) \, .
\end{eqnarray}
Moreover, Eq. (\ref{Cons}) means that $\Omega_{\Sigma}$ is conserved in
$\tau$-evolution. The two-form (\ref{degene}) is degenerate in the sense that
\begin{eqnarray}
\left ( \Omega_{\Sigma} \right ) _{AB} v^B =0 \, ,
\end{eqnarray}
for a non-trivial vector field $v^B$ on $\Sigma$. Solving last equation one
finds that there is a single null direction, as expected because there is a
single first class constraint, given by
\begin{eqnarray}
v & = &  a(q,p,t) \left ( \frac{\partial H}{\partial p_i}
\frac{\partial}{\partial q^i} - \frac{\partial H}{\partial q^i}
\frac{\partial}{\partial p_i} + \frac{\partial }{\partial t} \right ) \, ,
\label{Nvector}
\end{eqnarray}
where $a(q,p,t)$ is an arbitrary non-vanishing function on $\Sigma$. However,
$\gamma$ has associated the Hamiltonian vector field $X_{d \gamma}$ on
$\Gamma_{ext}$, $(\Omega_{ext}) \cdot X_{d \gamma} = d \gamma$, where the dot
`$\cdot$' stands for contraction, and it is given by
\begin{eqnarray}
X_{d \gamma} & = &  \frac{\partial H}{\partial p_i} \frac{\partial}{\partial
q^i} - \frac{\partial H}{\partial q^i} \frac{\partial}{\partial p_i} +
\frac{\partial }{\partial t} \, . \label{NvectorII}
\end{eqnarray}
$X_{d\gamma}$ is globally defined on $\Gamma_{ext}$ and, in particular, on
$\Sigma$. From the condition
\begin{eqnarray}
i_{\ast} v = X_{d \gamma} \, ,
\end{eqnarray}
it follows that $a(q,p,t)=1$. So,
\begin{eqnarray}
v & = &  \frac{\partial H}{\partial p_i} \frac{\partial}{\partial q^i} -
\frac{\partial H}{\partial q^i} \frac{\partial}{\partial p_i} + \frac{\partial
}{\partial t} \, . \label{Nvector3}
\end{eqnarray}

{\it Definition}. The couple $(\Sigma, \Omega_{\Sigma})$ is the {\it
presymplectic phase space} for the parametrized Hamiltonian system. The name
presymplectic comes from the fact that the structure of Eq. (\ref{degene}) is
closed, degenerate, and defined on the odd-dimensional manifold $\Sigma$.

Note that the presymplectic phase space $(\Sigma, \Omega_{\Sigma})$ is a
well-defined structure even though $H$ might not explicitly depend on the time
variable $t$. Moreover, Eq. (\ref{degene}) can be written as\cite{Kastrup}
\begin{eqnarray}
\Omega_{\Sigma} & = & {\tilde\delta} p_i \wedge {\tilde\delta} q^i -
{\tilde\delta} H(q,p,t) \wedge {\tilde\delta} t \, , \nonumber\\
& = & {\tilde\delta} p_i \wedge {\tilde\delta}  q^i  - \left ( \frac{\partial
H }{\partial q^i} {\tilde\delta} q^i + \frac{\partial H}{\partial p_i}
{\tilde\delta} p_i + \frac{\partial H}{\partial t} {\tilde\delta} t \right )
\wedge {\tilde\delta} t \, , \nonumber\\
& = & \left ( {\tilde\delta} p_i + \frac{\partial H}{\partial q^i}
{\tilde\delta} t \right ) \wedge \left ( {\tilde\delta} q^i - \frac{\partial H
}{\partial p_i}{\tilde\delta} t \right )\, , \nonumber\\
& = & \alpha_i \wedge \beta^i \, ,
\end{eqnarray}
with
\begin{eqnarray}
\alpha_i & := & {\tilde\delta} p_i + \frac{\partial H}{\partial q^i}
{\tilde\delta} t \, , \nonumber\\
\beta^i & := & {\tilde\delta} q^i - \frac{\partial H }{\partial
p_i}{\tilde\delta} t \, .
\end{eqnarray}

So far, only the symplectic $(\Gamma_{ext},\Omega_{ext})$ and the
presymplectic  $(\Sigma, \Omega_{\Sigma})$ phase spaces have been analyzed
using the procedure of Ref. \cite{Witten}. Now, it will be studied the
so-called physical or reduced phase space. To do this, it will be convenient
to analyze first the issue of Dirac observables.

{\it Definition}. The {\it Dirac} or {\it physical observables} ${\cal O}$ for
the system are real functions on $\Sigma$, ${\cal O}:\Sigma\longrightarrow R$,
killed by the null vector $v$ of Eq. (\ref{Nvector3})
\begin{eqnarray}
v {\cal O} & = & \left ( \frac{\partial H}{\partial p_i}
\frac{\partial}{\partial q^i} - \frac{\partial H}{\partial q^i}
\frac{\partial}{\partial p_i} + \frac{\partial }{\partial t} \right ) {\cal O}
=0 \, . \label{DiracObs}
\end{eqnarray}
This means that the ${\cal O}$'s are constant along the orbits on $\Sigma$ to
which $v$ is tangent. Equation (\ref{DiracObs}) can be written as
\begin{eqnarray}
\left \{ {\cal O} , H \right \} + \frac{\partial {\cal O}}{\partial t} & = & 0
\Longleftrightarrow \frac{d {\cal O}}{dt}=0 \, . \label{CMot}
\end{eqnarray}
Thus, we have got the following:

{\bf result 1}. Dirac {\it observables} or {\it physical observables} ${\cal
O}$ are the {\it constants of motion with respect to} $t$ of the standard
(i.e., non-parametrized) Hamiltonian system because the bracket in Eq.
(\ref{CMot}) is the standard Poisson bracket on $\Gamma$ given by Eq.
(\ref{PoissDeff}).

Moreover, due to the fact $\Sigma$ is a (2n+1)-dimensional manifold and one
has a single linear differential equation on it for the unknowns ${\cal O}$,
then Eq. (\ref{DiracObs}) [or, equivalently, Eq. (\ref{CMot})] has $2n$
independent solutions, say, $({\cal Q}^i (q,p,t) , {\cal P}_i (q,p,t))$. In
addition, $({\cal Q}^i (q,p,t) , {\cal P}_i (q,p,t))$ can be chosen to be
canonical (again, with respect to the bracket of Eq. (\ref{PoissDeff}))
\begin{eqnarray}
\{ {\cal Q}^i (q,p,t) , {\cal Q}^j (q,p,t) \} & = & 0 \, , \nonumber\\
\{ {\cal Q}^i (q,p,t) , {\cal P}_j (q,p,t) \} & = & \delta^i_j \, , \nonumber\\
\{ {\cal P}_i (q,p,t), {\cal P}_j (q,p,t) \} & = & 0 \, . \label{RedVar}
\end{eqnarray}

The specification of the physical observables $({\cal Q}^i (q,p,t), {\cal P}_i
(q,p,t) )$ to the independent concrete values $(Q^i,P_i)$ they can have
\begin{eqnarray}
Q^i & = & {\cal Q}^i (q,p,t) \, , \nonumber\\
P_i & = & {\cal P}_i (q,p,t) \, , \label{NEW}
\end{eqnarray}
gives rise to curves in $\Sigma$ (one curve for each value of $(Q^i,P_i)$),
which are called {\it motions}, {\it orbits}, {\it stories}, or {\it physical
states}. These curves are precisely those to which $v$ in Eq. (\ref{Nvector3})
is tangent.

{\it Definition}. The {\it reduced} or {\it physical phase space} for the
parametrized Hamiltonian system is the couple $(\Gamma_{phys},
\Omega_{phys})$, where the points of the $2n$-dimensional manifold
$\Gamma_{phys}$ represent all the possible motions for the system; each point
being (locally) labelled by the independent coordinates $(Q^i,P_i)$, and the
symplectic structure $\Omega_{phys}$ (written in these coordinates) is defined
by
\begin{eqnarray}
\Omega_{phys} & := & {\tilde\delta} P_i \wedge {\tilde\delta} Q^i \, .
\label{SymRed}
\end{eqnarray}

It is clear that the locus of Eq. (\ref{NEW}) defines a projection map $\pi$
from the presymplectic phase space $\Sigma$ to the physical phase space
$\Gamma_{phys}$ \cite{MonGRG}
\begin{eqnarray}
\pi & : & \Sigma \longrightarrow \Gamma_{phys}\, , \nonumber\\
& & {\bf p} \longmapsto {\bf q}= \pi({\bf p}) \, ,
\end{eqnarray}
that sends a point ${\bf p}$ on $\Sigma$ to the orbit ${\bf q}$ to which it
belongs. More precisely, let $(Q^i, P_i)$ be local coordinates around the
point ${\bf q}= \pi({\bf p})$
\begin{eqnarray}
\left ( Q^i (\pi({\bf p})), P_i (\pi({\bf p}) ) \right ) & = & \left ( (Q^i
\circ
\pi ) ({\bf p}) , (P_i \circ \pi)({\bf p}) \right ) \, , \nonumber\\
& = & \left (  \left ( \pi^{\ast} Q^i \right ) ({\bf p}), \left ( \pi^{\ast}
P_i \right ) ({\bf p}) \right )
\, , \nonumber\\
& = & \left ( {\cal Q}^i ({\bf p}), {\cal P}_i ({\bf p}) \right ) \, ,
\end{eqnarray}
i.e.,
\begin{eqnarray}
\pi^{\ast} Q^i & = & {\cal Q}^i \, , \nonumber\\
\pi^{\ast} P_i & = & {\cal P}_i \, .
\end{eqnarray}

Taking the map $\pi$ into account, the pull-back of various geometrical
objects on $\Gamma_{phys}$ to $\Sigma$ can be computed. In particular, the
pull-back of the one-forms ${\tilde\delta} Q^i$ and ${\tilde\delta} P_i$ on
$\Gamma_{phys}$ to $\Sigma$ are
\begin{eqnarray}
\pi^{\ast} {\tilde\delta} Q^i & = & {\tilde \delta} \left ( \pi^{\ast} Q^i
\right ) =
 {\tilde\delta} {\cal Q}^i \, , \nonumber\\
& = & \frac{\partial {\cal Q}^i}{\partial q^j} {\tilde\delta} q^j +
\frac{\partial {\cal Q}^i}{\partial p_j} {\tilde\delta} p_j + \frac{\partial
{\cal Q}^i}{\partial t} {\tilde\delta} t \, , \label{pull1}\\
& = & \frac{\partial {\cal Q}^i}{\partial q^j} {\tilde\delta} q^j +
\frac{\partial {\cal Q}^i}{\partial p_j} {\tilde\delta} p_j - \{ {\cal Q}^i ,
H \} \,
{\tilde\delta} t \, , \nonumber\\
& = & \frac{\partial {\cal Q}^i}{\partial q^j} \left ( {\tilde\delta} q^j -
\frac{\partial H }{\partial p_j}{\tilde\delta} t \right ) + \frac{\partial
{\cal Q}^i}{\partial p_j} \left ( {\tilde\delta} p_j + \frac{\partial
H}{\partial
q^j} {\tilde\delta} t  \right ) \, , \nonumber\\
& = & \frac{\partial {\cal Q}^i}{\partial q^j} \beta^j + \frac{\partial {\cal
Q}^i}{\partial
p_j} \alpha_j \, , \label{pull2}\\
\pi^{\ast} {\tilde\delta} P_i & = & {\tilde \delta} \left ( \pi^{\ast} P_i
\right ) = {\tilde\delta} {\cal P}_i \, , \nonumber\\
& = & \frac{\partial {\cal P}_i}{\partial q^j} {\tilde\delta} q^j +
\frac{\partial {\cal P}_i}{\partial p_j} {\tilde\delta} p_j + \frac{\partial
{\cal P}_i}{\partial t} {\tilde\delta} t \, , \label{pull3}\\
& = & \frac{\partial {\cal P}_i}{\partial q^j} {\tilde\delta} q^j +
\frac{\partial {\cal P}_i}{\partial p_j} {\tilde\delta} p_j - \{ {\cal P}_i ,
H \} \,
{\tilde\delta} t \, , \nonumber\\
& = & \frac{\partial {\cal P}_i}{\partial q^j} \left ( {\tilde\delta} q^j -
\frac{\partial H }{\partial p_j}{\tilde\delta} t \right ) + \frac{\partial
{\cal P}_i}{\partial p_j} \left ( {\tilde\delta} p_j + \frac{\partial
H}{\partial
q^j} {\tilde\delta} t  \right ) \, , \nonumber\\
& = & \frac{\partial {\cal P}_i}{\partial q^j} \beta^j + \frac{\partial {\cal
P}_i}{\partial p_j} \alpha_j \, , \label{pull4}
\end{eqnarray}
where Eq. (\ref{CMot}) was used. Thus, the right side of Eq. (\ref{pull1})
(equivalently, the right side of Eq. (\ref{pull2})) is the pull-back of
${\tilde\delta} Q^i$ to $\Sigma$. In addition, the right side of Eq.
(\ref{pull3}) (equivalently, the right side of Eq. (\ref{pull4})) is the
pull-back of ${\tilde\delta} P_i$ to $\Sigma$.

Using these results, the pull-back of the symplectic structure $\Omega_{phys}$
on $\Gamma_{phys}$ to the constraint surface $\Sigma$ can be obtained. The
resulting two-form on $\Sigma$ is precisely the presymplectic structure
$\Omega_{\Sigma}$
\begin{eqnarray}
\pi^{\ast} \Omega_{phys} & = & \Omega_{\Sigma} \, .
\end{eqnarray}
{\it Proof}: using Eqs. (\ref{SymRed}), (\ref{pull2}) and (\ref{pull4})
\begin{eqnarray}
\pi^{\ast} \Omega_{phys} & = & \left ( \pi^{\ast} {\tilde\delta} P_i \right )
\wedge
\left ( \pi^{\ast} {\tilde\delta} Q^i \right ) \, , \nonumber\\
& = & \frac{1}{2}  \left ( \frac{\partial {\cal P}_i}{\partial q^k}
\frac{\partial {\cal Q}^i}{\partial q^j} - \frac{\partial {\cal P}_i}{\partial
q^j} \frac{\partial {\cal Q}^i}{\partial q^k} \right ) \beta^k \wedge \beta^j
+ \frac12 \left ( \frac{\partial {\cal P}_i}{\partial p_k} \frac{\partial
{\cal Q}^i}{\partial p_j}  - \frac{\partial {\cal P}_i}{\partial p_j}
\frac{\partial {\cal Q}^i}{\partial p_k} \right ) \alpha_k \wedge \alpha_j
\nonumber\\
& & + \left ( \frac{\partial {\cal P}_i}{\partial p_k} \frac{\partial {\cal
Q}^i}{\partial q^j}- \frac{\partial {\cal P}_i}{\partial q^j} \frac{\partial
{\cal Q}^i}{\partial p_k} \right ) \alpha_k \wedge \beta^j
\, , \nonumber\\
& = & \alpha_i \wedge \beta^i \, , \nonumber\\
& = & \Omega_{\Sigma} \, ,
\end{eqnarray}
because of Eq. (\ref{RedVar}) [cf. Refs. \cite{RovPO,RovNote,RovNote3}].

Note that the {\it meaning and the range} of the variables $(Q^i , P_i)$ do
{\it not} correspond, in the generic case, with the ones of the variables
$(q^i , p_i)$ at the initial time $t_0$, given by $(q^i_0 , p_{i0})$. Of
course, in particular $(Q^i,P_i)$ can be taken to be $(q^i_0 , p_{i0})$.

In summary, the various geometrical structures involved in the dynamics of
parametrized Hamiltonian systems are: $(\Gamma_{ext}, \Omega_{ext})$,
$(\Sigma, \Omega_{\Sigma})$, $(\Gamma_{phys}, \Omega_{phys})$, and
$(\Gamma,\omega)$. In particular, the physical observables for the system
introduces $(\Gamma,\omega)$ again in the formalism of the parametrized
Hamiltonian system and this allows us to get the physical phase space for the
system $(\Gamma_{phys},\Omega_{phys})$. The interplay between these structures
is shown in the next diagram
\begin{eqnarray}
T^{\ast} {\cal C}_{ext} \stackrel{i}{\longleftarrow} \Sigma
\stackrel{\pi}{\longrightarrow} \Gamma_{phys} \, ,
\end{eqnarray}
or, in a more familiar notation
\begin{eqnarray}
\Gamma \times R^2 \stackrel{i}{\longleftarrow} \Gamma \times R
\stackrel{\pi}{\longrightarrow} \Gamma_{phys} \, ,
\end{eqnarray}
or, equivalently
\begin{eqnarray}
T^{\ast} {\cal C} \times R^2 \stackrel{i}{\longleftarrow} T^{\ast} {\cal C}
\times R \stackrel{\pi}{\longrightarrow} \Gamma_{phys} \, ,
\end{eqnarray}
[{\it cf.} Refs. \cite{Witten,Lee,Barnich}].

{\it Relationship with the canonical formalism a la Dirac}. In the canonical
formalism {\it a la} Dirac it is possible to get the physical phase space by
fixing the gauge freedom. In the present case, a single gauge condition would
be enough because there is a single first class constraint. In general, by
choosing a gauge condition, Dirac's method leads to canonical coordinates to
label the points of $\Gamma_{phys}$. On the other hand, in the present
approach no gauge condition has been chosen to reach $\Gamma_{phys}$. What is
then the relationship of the present approach to the fact of choosing a gauge
condition in Dirac's method to reach $\Gamma_{phys}$? Well, the answer is as
follows: in the present approach there still exists the freedom to choose the
{\it particular} set of Dirac observables that form $({\cal Q}^i (q,p,t) ,
{\cal P}_i (q,p,t))$, i.e., there exists the freedom to choose the canonical
coordinates on $\Gamma_{phys}$. This freedom corresponds, precisely, to the
freedom of picking a gauge condition in Dirac's method.

\section{Freedom in the choice of the symplectic structure $\Omega_{ext}$ in
$\Gamma_{ext}$ keeping the same constraint surface $\Sigma$} Note that a more
covariant description of $(\Gamma_{phys}, \Omega_{phys})$ (and therefore of
$(\Gamma_{ext}, \Omega_{ext})$) would be given by labelling the points of
$\Gamma_{phys}$ with {\it arbitrary} labels $X^a$ (not necessarily canonical
ones) with the only restriction that the independent coordinates $X^a$ label
physically distinct states. In these coordinates, $\Omega_{phys}$ would have
the form
\begin{eqnarray}
\Omega_{phys} & = & \frac{1}{2} (\Omega_{phys})_{ab} (X) \, d X^a \wedge d X^b
\,
, \nonumber\\
\{ X^a , X^b  \}_{phys} & := & \Omega^{ab}_{phys} (X) \, . \label{NonCan}
\end{eqnarray}
To illustrate this point consider the action principle (\ref{HamActPar}) with
$(x^{\mu})=(x,y,t,p_x,p_y,p_t)$ and
\begin{eqnarray}
G_0 := p_t + \frac{1}{2} \left ( \frac{p^2_x}{m} + m \omega^2 x^2 +
\frac{p^2_y}{m} + m \omega^2 y^2 \right ) \, ,
\end{eqnarray}
which yields the equations of motion
\begin{eqnarray}\label{GerMendEM}
& & {\dot x} = \lambda \frac{p_x}{m}\, , \quad {\dot y} = \lambda
\frac{p_y}{m}\, , \quad {\dot t} = \lambda \, , \nonumber\\
& &  {\dot p}_x = - \lambda m \omega^2 x\, , \quad {\dot p}_y = - \lambda m
\omega^2 y \, , \quad {\dot p}_t =0 \, ,
\end{eqnarray}
and the constraint
\begin{eqnarray}
G_0 \approx 0 \, . \label{sameConst}
\end{eqnarray}
As it was explained in Sec. II, the standard procedure consists in taking
$(x^{\mu})=(x,y,t,p_x,p_y,p_t)$ as canonical coordinates with the standard
symplectic structure on $\Gamma_{ext}$ given by
\begin{eqnarray}\label{usual}
\Omega_0 = d p_x \wedge dx + dp_y \wedge dy + d p_t \wedge dt \, ,
\end{eqnarray}
and the corresponding symplectic structure $\omega_0$ on $\Gamma_{phys}$ given
by
\begin{eqnarray}\label{usualII}
\omega_0 = d p_{x0} \wedge d x_0 + d p_{y0} \wedge d y_0 \, .
\end{eqnarray}
However, it is not necessary that the coordinates
$(x^{\mu})=(x,y,t,p_x,p_y,p_t)$ which label the points of $\Gamma_{ext}$ are
canonical ones. It is possible, for instance, to choose\footnote{Note that we
are {\it not} making a change of coordinates, we are working with the {\it
same} set of coordinates, i.e., what changes is the symplectic structure.}
\begin{eqnarray}\label{GerMend}
\Omega & = & \Lambda m^2 \omega^2 x y d x \wedge d y - d x \wedge d p_x
+ \Lambda x p_y dx \wedge d p_y \nonumber\\
& & - \Lambda y p_x d y \wedge d p_x -  dy \wedge d p_y + \Lambda \frac{p_x
p_y }{m^2 \omega^2}d p_x \wedge d p_y + d p_t \wedge dt \, ,
\end{eqnarray}
where $\Lambda$ is an arbitrary real constant, as the symplectic structure on
$\Gamma_{ext}$ with the corresponding symplectic structure\cite{GFTorres}
\begin{eqnarray}
\omega & = & \Lambda m^2 \omega^2 x_0 y_0 d x_0 \wedge d y_0 - d x_0 \wedge d
p_{x0} + \Lambda x_0 p_{y0} d x_0 \wedge d p_{y0} \nonumber\\
& & - \Lambda y_0 p_{x0} d y_0 \wedge d p_{x0} -  d y_0 \wedge d p_{y0} +
\Lambda \frac{p_{x0} p_{y0} }{m^2 \omega^2}d p_{x0} \wedge d p_{y0}\, ,
\end{eqnarray}
or, equivalently,
\begin{eqnarray}
\{ x_0 , y_0 \} = \Lambda \frac{p_{x0} p_{y0}}{m^2 \omega^2}\, , \quad \{ x_0
, p_{x0} \} = 1\, , \quad \{ x_0 , p_{y0} \} = - \Lambda y_0 p_{x0}\, ,
\nonumber\\
\{ y_0 , p_{x0} \} = \Lambda x_0 p_{y0} \, ,\quad \{ y_0 , p_{y0} \} = 1 \, ,
\quad \{ p_{x0} , p_{y0} \} = \Lambda m^2 \omega^2 x_0 y_0 \, ,
\end{eqnarray}
on $\Gamma_{phys}$ and keeping the same constraint (\ref{sameConst}). In fact,
note that the equations
\begin{eqnarray}
\frac{d x^{\mu}}{d \tau} = \{ x^{\mu} , \lambda G_0 \} = \lambda
\Omega^{\mu\nu} \frac{\partial G_0 }{\partial x^{\nu}} \, ,
\end{eqnarray}
where $(\Omega^{\mu\nu})$ is the inverse matrix of the symplectic matrix given
in (\ref{GerMend}), exactly reproduce the equations of motion of Eq.
(\ref{GerMendEM}). Moreover, note that the presymplectic two-forms $i^{\ast}
\Omega_0$ and $i^{\ast} \Omega$, defined on $\Sigma$, have the same null
vector given in Eq. (\ref{NULL}). Furthermore, the gauge transformation on the
variables $(x^{\mu})=(x,y,t,p_x,p_y,p_t)$ computed with the standard
symplectic structure (\ref{usual}) is exactly the same one computed with the
symplectic structure given in Eq. (\ref{GerMend}) [see also Secs. IV, V, and
VI]. These two choices can be represented in the following diagram:
\begin{eqnarray}
(\Gamma_{ext}, \Omega_0) \stackrel{i}{\longleftarrow} (\Sigma, i^{\ast}
\Omega_0) \stackrel{\pi}{\longrightarrow} (\Gamma_{phys}, \omega_0) \,
, \\
(\Gamma_{ext}, \Omega) \stackrel{i}{\longleftarrow} (\Sigma, i^{\ast} \Omega)
\stackrel{\pi}{\longrightarrow} (\Gamma_{phys}, \omega)\, ,
\end{eqnarray}
respectively. Thus, we have got the following:

{\bf result 2}. The specification of both the constraint (surface) $\gamma
\approx 0$ and the dynamical equations of motion [see Eqs. (\ref{GerMendEM})
and (\ref{sameConst})] do {\it not} uniquely fix the symplectic structure on
the extended phase space $\Gamma_{ext}$ and therefore do {\it not} uniquely
fix the presymplectic structure on the constraint surface $\Sigma$ and do {\it
not} uniquely fix the symplectic structure on the physical phase space
$\Gamma_{phys}$. In particular, there is not need of choosing the coordinates
that label the points of $\Gamma_{ext}$ and $\Gamma_{phys}$ to be canonical
coordinates, which is always the case in Dirac's approach, and there exists
the freedom to choose non-canonical symplectic structures on $\Gamma_{ext}$
and $\Gamma_{phys}$. Of course, there are many other ways of choosing the
symplectic structure on $\Gamma_{ext}$ and the corresponding presymplectic
structure on $\Sigma$ and the symplectic structure on $\Gamma_{phys}$ keeping
the same constraint (\ref{sameConst}). We have just listed two of these
possible choices.

Finally, note that by means of Darboux's theorem all of these non-canonical
expressions for the symplectic structure on $\Omega_{ext}$ (and therefore
non-canonical expressions for $\Omega_{phys}$) can (locally) acquire the
canonical form. However, it would be interesting not to write them in a
canonical form and to explore the consequences of these possible choices in
the quantum theory. In particular, to understand how this situation is handled
in the framework of Dirac's quantization as well as in the framework of
reduced phase space quantization.

\section{Freedom in the choice of both the symplectic structure $\Omega_{ext}$
in $\Gamma_{ext}$ and the constraint surface $\Sigma$} It is remarkable that
$H$ in Eq. (\ref{NonPar}) does not need to be the energy, even for
conservative systems. To see this, consider again the equations of motion for
the two-dimensional isotropic harmonic oscillator
\begin{eqnarray}
{\dot x} & = & \frac{p_x}{m}\, , \quad {\dot y} = \frac{p_y}{m}\, , \quad
{\dot p}_x = - m \omega^2 x\, , \quad {\dot p}_y = - m \omega^2 y \, ,
\label{HOEmotion}
\end{eqnarray}
where the dot `.' stands for the total derivative with respect to $t$. It is
well known that the equations of motion (\ref{HOEmotion}) can be obtained from
the action principle
\begin{eqnarray}\label{Standard}
S[x,y,p_x,p_y] & = & \int^{t_2}_{t_1} \left [ {\dot x} p_x + {\dot y } p_y -
H_0 \right ] dt \, , \nonumber\\
H_0 & = & \frac{1}{2} \left ( \frac{p^2_x}{m} + m \omega^2 x^2 +
\frac{p^2_y}{m} + m \omega^2 y^2 \right )\, .
\end{eqnarray}
However, the equations of motion (\ref{HOEmotion}) can also, for instance, be
obtained from the action principle\cite{GFTorres,GFTorres2}
\begin{eqnarray}\label{Gerardo}
S [x,y,p_x,p_y] & = & \int^{t_2}_{t_1} \left [ {\dot x} p_y + {\dot y} p_x -
J_1
\right ]dt \, , \nonumber\\
J_1 & = & \frac{p_x p_y}{m} + m \omega^2 xy \, .
\end{eqnarray}
Note that $H_0$ in Eq. (\ref{Standard}) {\it is} bounded from below while
$J_1$ in Eq. (\ref{Gerardo}) is {\it not}. In addition, in Eq.
(\ref{Standard}) the canonical momenta of $(x,y)$ are their corresponding
kinetic momenta while in Eq. (\ref{Gerardo}) the canonical momenta of $(x,y)$
are not their kinetic momenta \footnote{In classical field theory is a common
fact that the canonical momenta of fields are not their linear momenta
obtained, for instance, from the energy-momentum tensor. However, Eq.
(\ref{Gerardo}) shows that this property is also present in systems with a
finite number of degrees of freedom.}.

The action principles (\ref{Standard}) and (\ref{Gerardo}) can be parametrized
incorporating the variable $t$ as a configuration variable. By doing this one
gets
\begin{eqnarray}\label{StandPar}
S[x,y,t,p_x,p_y,p_t,\lambda] & = & \int^{\tau_2}_{\tau_1} \left [ {\dot x} p_x
+ {\dot
y } p_y  + {\dot t} p_t - \lambda G_0 \right ] d\tau \, , \nonumber\\
G_0 & := & p_t + H_0 \, ,
\end{eqnarray}
and
\begin{eqnarray}\label{GerPar}
S_1 [x,y,t,p_x,p_y,p_t,\lambda] & = & \int^{\tau_2}_{\tau_1} \left [ {\dot x}
p_y + {\dot
y } p_x  + {\dot t} p_t - \lambda G_1 \right ] d \tau \, , \nonumber\\
G_1 & := & p_t + J_1 \, ;
\end{eqnarray}
respectively.

{\it Extended phase space}. In the case of the action (\ref{GerPar}), the
extended phase space $(\Gamma_{ext}, \Omega_1)$ is such that
$(x,y,t,p_x,p_y,p_t)$ are independent coordinates for labelling the points of
$\Gamma_{ext}$ and the symplectic structure $\Omega_1$ on $\Gamma_{ext}$ is
given by
\begin{eqnarray}\label{twoformGer}
\Omega_1 & = & d p_y \wedge dx + d p_x \wedge dy + d p_t \wedge d t \, ,
\end{eqnarray}
or, equivalently, the nonvanishing Poisson brackets are
\begin{eqnarray}
\{ x, p_y \}_1 = 1 \, , \quad \{ y , p_x \}_1 = 1 \, , \quad \{ t , p_t \}_1
=1 \, ,
\end{eqnarray}
instead of $\Omega_0$, given in Eq. (\ref{usual}), or equivalently,
\begin{eqnarray}
\{ x , p_x \}_0 =1\, ,\quad \{ y , p_y \}_0 =1\, , \quad \{ t, p_t \}_0 =1\,
\end{eqnarray}
which corresponds to the action (\ref{StandPar}).

{\it Presymplectic phase space}. According to (\ref{GerPar}), the
presymplectic phase space $\Sigma_1$ is defined as $\Sigma_1 = \{
(x,y,t,p_x,p_y,p_t) \in \Gamma_{ext} \mid G_1 = p_t + J_1 =0 \}$. The
presymplectic two-form $\Omega_{\Sigma_1}$ on $\Sigma_1$, induced by
(\ref{twoformGer}), is given by
\begin{eqnarray}\label{AAA}
\Omega_{\Sigma_1} & = & i^{\ast} \Omega_1 \nonumber\\
& = & d p_y \wedge dx + d p_x \wedge dy - d J_1 \wedge d t
\nonumber\\
& = & \left ( d p_y + m \omega^2 y dt \right ) \wedge \left ( dx -
\frac{p_x}{m} dt \right ) + \left ( d p_x + m \omega^2 x dt \right ) \wedge
\left ( d y - \frac{p_y}{m} dt \right ) \, .
\end{eqnarray}
The null vector $v$ of $\Omega_{\Sigma_1}$ in Eq. (\ref{AAA}) is given by
\begin{eqnarray}
v & = & \frac{p_x}{m} \frac{\partial}{\partial x} + \frac{p_y}{m}
\frac{\partial}{\partial y} - m \omega^2 x \frac{\partial}{\partial p_x} - m
\omega^2 y \frac{\partial}{\partial p_y} + \frac{\partial}{\partial t} \, .
\label{NULL}
\end{eqnarray}
Note that this null vector is also the null vector of the standard symplectic
structure $\Omega_{\Sigma_0}$
\begin{eqnarray}
\Omega_{\Sigma_0} & = & i^{\ast} \Omega_0 \nonumber\\
& = & d p_x \wedge d x + d p_y \wedge d y - d H_0 \wedge dt \, .
\end{eqnarray}
Therefore, we have got the following:

{\bf result 3}. There exists the freedom to choose both the symplectic
structure $\Omega_{ext}$ on $\Gamma_{ext}$ and the equation that defines the
constraint surface $\Sigma$ with the only restriction that these two choices
combine in such a way that the null vector $v$ is the same for any choice of
the pair $(\Omega_{ext}, \Sigma)$. Note that the integral curves to which $v$
is tangent belong to $\Sigma_0$ and also to $\Sigma_1$ (and also to any other
constraint surface $\Sigma_2$ such that $(\Omega_2, \Sigma_2)$ generates the
same null vector, see the end part of this section). Therefore, the integral
curves are also integral curves of the intersection of all these surfaces
$\Sigma_0$, $\Sigma_1$, $\Sigma_2$, etc.

{\bf result 4}. Due to the fact that the null vector $v$ is the same for any
choice and because the physical observables ${\cal O}$ for the system are
those functions on $\Sigma$ such that $v {\cal O}=0$ then Dirac observables
${\cal O}$ are the same for any choice of the couple $(\Omega_{ext}, \Sigma)$.
Moreover, in Sec.~\ref{Geometry} it has been shown that Dirac observables
${\cal O}$ are the constants of motion with respect to $t$ of the
unparametrized system. It is important to recall that to be a constant of
motion ${\cal O}$ is just a property of the equations of motion, and does not
depend on the choice of the Hamiltonian or of the symplectic structure
\cite{GFTorres,GFTorres2}. This explains why Dirac observables ${\cal O}$ are
{\it independent} of the choice of the symplectic structure $\Omega_{ext}$ on
$\Gamma_{ext}$ and of the specification of $\Sigma$ in the sense explained
above.

{\it Physical phase space}. The points of $\Gamma_{phys}$ can be labelled with
the independent coordinates $(x_0,y_0,p_{x0},p_{y0})$ and the symplectic
two-form on it is
\begin{eqnarray}
\omega_1 := d p_{y0} \wedge d x_0 + d p_{x0} \wedge d y_0 \, ,
\end{eqnarray}
or, equivalently, the nonvanishing Poisson brackets are
\begin{eqnarray}
\{ x_0 , p_{y0} \} & = & 1 \, , \quad \{ y_0 , p_{x0} \} = 1 \, ,
\end{eqnarray}
in opposition to $\omega_0$, given in Eq. (\ref{usualII}), which corresponds
to (\ref{StandPar}).

In summary, from (\ref{GerPar}) one has
\begin{eqnarray}
(\Gamma_{ext}, \Omega_1 ) \stackrel{i}{\longleftarrow} (\Sigma_1 ,
\Omega_{\Sigma_1}) \stackrel{\pi}{\longrightarrow} (\Gamma_{phys}, \omega_1)\,
,
\end{eqnarray}
while from (\ref{StandPar}) one has
\begin{eqnarray}
(\Gamma_{ext}, \Omega_0) \stackrel{i}{\longleftarrow} (\Sigma_0 ,
\Omega_{\Sigma_0}) \stackrel{\pi}{\longrightarrow} (\Gamma_{phys}, \omega_0)\,
.
\end{eqnarray}

There are many other ways of choosing the pair $(\Omega_{ext},\Sigma)$ or,
equivalently, the pair $(\Omega_{ext}, G)$ where $G$ is the first class
constraint. We just list other two of these pairs:

1)
\begin{eqnarray}
(\Gamma_{ext}, \Omega_2) \stackrel{i}{\longleftarrow} (\Sigma_2 ,
\Omega_{\Sigma_2}) \stackrel{\pi}{\longrightarrow} (\Gamma_{phys}, \omega_2)\,
,
\end{eqnarray}
with
\begin{eqnarray}\label{S2}
\Omega_2 & := & d x \wedge d p_x - d y \wedge d p_y + d p_t \wedge dt \, ,
\nonumber\\
\Omega_{\Sigma_2} & = & d x \wedge d p_x - d y \wedge d p_y - d J_2 \wedge dt
\nonumber\\
& = & d x \wedge d p_x - d y \wedge d p_y + m \omega^2 x dx \wedge dt - m
\omega^2 y dy \wedge dt \nonumber\\
& & + \frac{p_x}{m} d p_x \wedge dt - \frac{p_y}{m} d p_y \wedge dt \, ,
\nonumber\\
\omega_2 & = & d x_0 \wedge d p_{x0} - d y_0 \wedge d p_{y0}\, ,
\end{eqnarray}
because of
\begin{eqnarray}\label{C2}
 G_2
& := & p_t + J_2 \approx 0 \, , \quad J_2 = \frac{p^2_y - p^2_x}{2m} + \frac12
m \omega^2 (y^2 - x^2)\, .
\end{eqnarray}

2)
\begin{eqnarray}
(\Gamma_{ext}, \Omega_3) \stackrel{i}{\longleftarrow} (\Sigma_3 ,
\Omega_{\Sigma_3}) \stackrel{\pi}{\longrightarrow} (\Gamma_{phys}, \omega_3)\,
,
\end{eqnarray}
with
\begin{eqnarray}\label{S3}
\Omega_3 & := & m \omega d x \wedge dy + \frac{1}{m\omega} d p_x \wedge d p_y
+ d p_t \wedge dt \, , \nonumber\\
\Omega_{\Sigma_3} & = & m \omega d x \wedge dy + \frac{1}{m\omega} d p_x
\wedge d p_y - d J_3 \wedge dt  \nonumber\\
& = & m \omega d x \wedge dy + \frac{1}{m\omega} d p_x \wedge d p_y - \omega
p_y d x \wedge dt \nonumber\\
& & - \omega x d p_y \wedge dt + \omega p_x dy \wedge dt + \omega y d p_x
\wedge dt \, , \nonumber\\
\omega_3 & = & m \omega d x_0 \wedge d y_0 + \frac{1}{m\omega} d p_{x0} \wedge
d p_{y0}\, ,
\end{eqnarray}
because of
\begin{eqnarray}\label{C3}
G_3 & := & p_t + J_3 \approx 0 \, , \quad J_3 = \omega (x p_y - y p_x) \, .
\end{eqnarray}

Note that neither $J_2$ nor $J_3$ are bounded from below. We have seen that
the vector (\ref{NULL}) is the null vector of $\Omega_{\Sigma_0}$ and
$\Omega_{\Sigma_1}$. In addition, the vector (\ref{NULL}) is also the null
vector of $\Omega_{\Sigma_2}$ and $\Omega_{\Sigma_3}$. Due to the fact
$(x,y,p_x,p_y,t)$ are local coordinates for $\Sigma_{\mu}$, $\mu=0,1,2,3$, one
would say that all the surfaces $\Sigma_{\mu}$ are diffeomorphic.

Finally, note that even though the symplectic structures $\Omega_1$,
$\Omega_2$, and $\Omega_3$ (after re-scaling the coordinates) have the
canonical form specified by Darboux's theorem, they are distinct to the usual
canonical form $\{\, ,\}_0$ given by Eq. (\ref{usual}).

\section{Covariant description of gauge transformations}
The infinitesimal gauge transformation of the variables $(x,y,t,p_x,p_y,p_t)$
(which label the points of $\Gamma_{ext}$) generated by the constraint $G_0 =
p_t + H_0 \approx 0$ in (\ref{StandPar}) and the symplectic structure of Eq.
(\ref{usual}) is
\begin{eqnarray}
\delta_{\varepsilon} x & = & \varepsilon \{ x , G_0 \}_0 = \varepsilon
\frac{p_x}{m} \{ x , p_x \}_0 =
\varepsilon \frac{p_x}{m} \, , \nonumber\\
\delta_{\varepsilon} y & = & \varepsilon \{ y , G_0 \}_0 = \varepsilon
\frac{p_y}{m} \{ y , p_y \}_0
= \varepsilon \frac{p_y}{m} \, , \nonumber\\
\delta_{\varepsilon} t & = & \varepsilon \{ t , G_0 \}_0 = \varepsilon \{ t ,
p_t \}_0 = \varepsilon \, ,
\nonumber\\
\delta_{\varepsilon} p_x & = & \varepsilon \{ p_x , G_0 \}_0 = \varepsilon m
\omega^2 x \{ p_x , x \}_0
= -\varepsilon m \omega^2 x \, , \nonumber\\
\delta_{\varepsilon} p_y & = & \varepsilon \{ p_y , G_0 \}_0 = \varepsilon m
\omega^2 y \{ p_y , y \}_0
= -\varepsilon m \omega^2 y \, , \nonumber\\
\delta_{\varepsilon} p_t & = & \varepsilon \{ p_t , G_0 \}_0 = 0 \, .
\label{StandGT}
\end{eqnarray}

On the other hand, the infinitesimal gauge transformation of the variables
$(x,y,t,p_x,p_y,p_t)$ generated by the constraint $G_1 = p_t + J_1 \approx 0$
in (\ref{GerPar}) and the symplectic structure $\Omega_1$ of Eq.
(\ref{twoformGer}) is
\begin{eqnarray}
\delta_{\varepsilon} x & = & \varepsilon \{ x, G_1 \}_1 =  \varepsilon
\frac{p_x}{m}\{ x , p_y \}_1 =
\varepsilon \frac{p_x}{m} \, , \nonumber\\
\delta_{\varepsilon} y & = & \varepsilon \{ y , G_1 \}_1 = \varepsilon
\frac{p_y}{m} \{ y , p_x \}_1
= \varepsilon \frac{p_y}{m} \, , \nonumber\\
\delta_{\varepsilon} t & = & \varepsilon \{ t , G_1 \}_1 = \varepsilon \{ t ,
p_t \}_1 = \varepsilon \, ,
\nonumber\\
\delta_{\varepsilon} p_x & = & \varepsilon \{ p_x , G_1 \}_1 = \varepsilon m
\omega^2 x \{ p_x , y \}_1 =
 - \varepsilon m \omega^2 x \, , \nonumber\\
\delta_{\varepsilon} p_y & = & \varepsilon \{ p_y , G_1 \}_1 = \varepsilon m
\omega^2 y \{ p_y , x \}_1
= -\varepsilon m \omega^2 y \, , \nonumber\\
\delta_{\varepsilon} p_t & = & \varepsilon \{ p_t , G_1 \}_1 = 0 \, .
\label{GerGT}
\end{eqnarray}

In the same way, the infinitesimal gauge transformation generated by the
constraint $G_2 = p_t + J_2 \approx 0$ and the symplectic structure $\Omega_2$
of Eq. (\ref{S2}) is
\begin{eqnarray}
\delta_{\varepsilon} x & = & \varepsilon \{ x , G_2 \}_2 = - \varepsilon
\frac{p_x}{m}\{ x , p_x \}_2 =
\varepsilon \frac{p_x}{m} \, , \nonumber\\
\delta_{\varepsilon} y & = & \varepsilon \{ y , G_2 \}_2 = \varepsilon
\frac{p_y}{m} \{ y , p_y \}_2 = \varepsilon \frac{p_y}{m}\, , \nonumber\\
\delta_{\varepsilon} t & = & \varepsilon \{ t, G_2 \}_2 = \varepsilon \{ t ,
p_t \}_2 = \varepsilon \, , \nonumber\\
\delta_{\varepsilon} p_x & = & \varepsilon \{ p_x , G_2 \}_2 = - \varepsilon m
\omega^2 x \{ p_x , x \}_2 = - \varepsilon m \omega^2 x \, , \nonumber\\
\delta_{\varepsilon} p_y & = & \varepsilon \{ p_y , G_2 \}_2 = \varepsilon m
\omega^2 y \{ p_y , y \}_2 = - \varepsilon m \omega^2 y \, , \nonumber\\
\delta_{\varepsilon} p_t & = & \varepsilon \{ p_t , G_2 \}_2 = 0 \,
.\label{S2GT}
\end{eqnarray}

Finally, the infinitesimal gauge transformation generated by the constraint
$G_3 = p_t + J_3 \approx 0$ and the symplectic structure $\Omega_3$ of Eq.
(\ref{S3}) is
\begin{eqnarray}\label{S3GT}
\delta_{\varepsilon} x & = & \varepsilon \{ x, G_3 \}_3 = -\varepsilon \omega
p_x \{ x, y \}_3 = \varepsilon \frac{p_x}{m}\, , \nonumber\\
\delta_{\varepsilon} y & = & \varepsilon \{ y , G_3 \}_3 =\varepsilon \omega
p_y \{ y , x \}_3 = \varepsilon \frac{p_y}{m} \, , \nonumber\\
\delta_{\varepsilon} t & = & \varepsilon \{ t , G_3 \}_3 =  \varepsilon \{ t ,
p_t \}_3 = \varepsilon \, , \nonumber\\
\delta_{\varepsilon} p_x & = & \varepsilon \{ p_x , G_3 \}_3 = \varepsilon
\omega x \{ p_x , p_y \}_3 =- \varepsilon m \omega^2 x \, , \nonumber\\
\delta_{\varepsilon} p_y & = & \varepsilon \{ p_y , G_3 \}_3 = -\varepsilon
\omega y \{ p_y , p_x \}_3 = - \varepsilon m \omega^2 y \, , \nonumber\\
\delta_{\varepsilon} p_t & = & \varepsilon \{ p_t , G_3 \}_3 = 0 \, .
\end{eqnarray}
Note that the right hand side of Eqs. (\ref{StandGT}), (\ref{GerGT}),
(\ref{S2GT}), and (\ref{S3GT}) are the same. We have got the following:

{\bf result 5}. Equations (\ref{StandGT}), (\ref{GerGT}), (\ref{S2GT}), and
(\ref{S3GT}) mean that the gauge transformation of the variables that label
the points of $\Gamma_{ext}$ is independent of the choice of the symplectic
structure $\Omega_{ext}$ on $\Gamma_{ext}$ and of the form of specifying
$\Sigma$ in the sense explained above. There is an easy way of understanding
the cause of this phenomenon. Assuming, for the moment, that there exists just
a single first class constraint $G_0 \approx 0$ (which is the case considered
so far in this paper) then the Hamiltonian formalism {\it a la} Dirac
\cite{Dirac} says that the gauge transformation on any function $F$ generated
by $G_0 \approx 0$ is
\begin{eqnarray}\label{DiracGT}
\delta_{\varepsilon} F = \varepsilon \{ F , G_0 \}_0 \, .
\end{eqnarray}
However, the right-hand side of Eq. (\ref{DiracGT}) can, instead of using the
pair $(\{ \ , \}_0, G_0)$ in Eq. (\ref{DiracGT}), be obtained from a new,
different, pair $(\{ \ , \}_{new}, G_{new} )$
\begin{eqnarray}\label{CovarGT}
\delta_{\varepsilon} F = \varepsilon \{ F , G_{new} \}_{new} \, ,
\end{eqnarray}
i.e., there exists an ambiguity, a freedom, in the choice of the symplectic
structure (or, equivalently, the Poisson brackets) and in the form that the
first class constraint is specified ($G_0 \approx 0$ or $G_{new}\approx 0$) in
such a way that any pair (of these choices) generates the same gauge
transformation on $F$ \footnote{Note that this ambiguity is not of the same
kind than the one that it is involved in the Abelianization of constraints
\cite{Henneaux}.}. In Dirac's approach one uses the usual canonical form of
the symplectic structure (equivalently, the canonical form of the Poisson
brackets,$\{ \ , \}_0$). However, we have seen that it is not mandatory to
choose $\{\, ,\}_0$ on $\Gamma_{ext}$ and that other choices are allowed, the
only restriction on these choices is that they generate the same gauge
transformation by choosing the appropriate form for the first class
constraint.

{\bf result 6}. Using the result 5 or, equivalently, Eqs. (\ref{StandGT}) and
(\ref{GerGT}) one has that {\it any} gauge-invariant function $O$ on
$\Gamma_{ext}$ (under the gauge transformation) has the same functional form
independent of the choice of the couple $(\Omega_0, G_0)$, $(\Omega_1, G_1)$,
$(\Omega_2, G_2)$, and $(\Omega_3, G_3)$ (and also of any other allowed
choices).

\section{The `algebra' of gauge-invariant functions on $\Gamma_{ext}$ depends on
the choice of $\Omega_{ext}$ and $\Sigma$}

{\bf result 7}. Even though gauge-invariant functions $O: \Gamma_{ext}
\rightarrow R$ have the same functional form independently of the choice of
the couple $(\Omega_0, G_0)$ or $(\Omega_1, G_1)$ (or any other allowed
choice), the Poisson brackets among the gauge-invariant functions $O$ on
$\Gamma_{ext}$ might or might not, in the generic case, form a Lie algebra
simply because the Lie algebra directly depends on the choice of the
symplectic structure on $\Gamma_{ext}$.

1) {\bf $su(2)$ algebra of observables}. The gauge-invariant functions on
$\Gamma_{ext}$
\begin{eqnarray}\label{GIF}
J_1 & = & \frac{p_x p_y}{m} + m \omega^2 xy \, , \nonumber\\
J_2 & = & \frac{p^2_y - p^2_x}{2m} + \frac{1}{2} m \omega^2 (y^2 - x^2) \, ,
\nonumber\\
J_3 & = & \omega (x p_y - y p_x) \, ,
\end{eqnarray}
satisfy\cite{GFTorres2}
\begin{eqnarray}\label{SU(2)}
\{ J_i , J_j \}_0 & = & 2 \omega \varepsilon_{ijk} J_k \, ,
\end{eqnarray}
with respect to the usual symplectic structure $\Omega_0$ (\ref{usual}) on
$\Gamma_{ext}$. It is clear that (\ref{SU(2)}) is a Lie algebra isomorphic to
the $su(2)$ (and $so(3)$) algebra.

2) {\bf $su(1,1)$ algebra of observables}. On the other hand, with respect to
the symplectic structure $\Omega_1$ (\ref{twoformGer}) the gauge-invariant
functions (\ref{GIF}) satisfy
\begin{eqnarray}
\{ J_1 , J_2 \}_1 & = & 0\, , \nonumber\\
\{ J_1 , J_3 \}_1 & = & 0 \, , \nonumber\\
\{ J_2 , J_3 \}_1 & = & - 2 \omega H_0 \, ,
\end{eqnarray}
which means that the gauge-invariant functions (\ref{GIF}) do {\it not} form a
Lie algebra with respect to the symplectic structure $\Omega_1$
(\ref{twoformGer}). Nevertheless, the set of gauge-invariant functions $\{
H_0, J_2, J_3 \}$ satisfy\cite{GFTorres2}
\begin{eqnarray}\label{noinv}
\{ H_0 , J_2 \}_1 & = & 2 \omega J_3 \, , \nonumber\\
\{ H_0 , J_3 \}_1 & = & - 2 \omega J_2 \, , \nonumber\\
\{ J_2 , J_3 \}_1 & = & - 2 \omega H_0 \, ,
\end{eqnarray}
with respect to the symplectic structure $\Omega_1$ of Eq. (\ref{twoformGer}),
which means that $\{ H_0, J_2 , J_3 \}$ generate an algebra isomorphic to
$su(1,1)$ \cite{GFTorres2}.

Note also that even though $H_0$ and $J_3$ are in involution with respect to
the symplectic structure (\ref{usual})
\begin{eqnarray}
\{ H_0 , J_3 \}_0 = 0 \, ,
\end{eqnarray}
they are not in involution with respect to the symplectic structure
(\ref{twoformGer}) because of the second line of Eq. (\ref{noinv}).

Finally, note also that
\begin{eqnarray}
J^2_1 + J^2_2 + J^2_3 = H^2_0 \, ,
\end{eqnarray}
holds independently of the choice of the symplectic structure, i.e., it does
not depend on (\ref{usual}) or (\ref{twoformGer}). However, the
gauge-invariant function that plays the role of Casimir directly depends on
the symplectic structure chosen.

\section{Systems with first class constraints only}
The results of the previous sections for parametrized nonrelativistic
Hamiltonian systems can be generalized to any generally covariant system with
first class constraints only. In the Hamiltonian formalism {\it a la} Dirac
one has
\begin{eqnarray}\label{XXMM}
\frac{d x^{\mu}}{d \tau} & = & \{ x^{\mu}, \lambda^a \gamma_a \}_0
\, , \nonumber\\
\gamma_a & \approx & 0 \, , \nonumber\\
\{ \gamma_a , \gamma_b \}_0 & = & C_{ab}\,^c \gamma_c \, ,
\end{eqnarray}
where $(x^{\mu})=(q^i,p_i)$, $\{ f, g \}_0 = \Omega^{\mu\nu}_0 \frac{\partial
f}{\partial x^{\mu}} \frac{\partial g}{\partial x^{\nu}}$, and $\{ x^{\mu},
x^{\nu} \}_0 = \Omega^{\mu\nu}_0 $ (which has the usual canonical form). If
one does not want to use this symplectic structure then the system
(\ref{XXMM}) can be replaced with
\begin{eqnarray}\label{CCC}
\frac{d x^{\mu}}{d\tau} & = & \{ x^{\mu}, \lambda^a G_a \} \, , \nonumber\\
G_a & \approx & 0 \, , \nonumber\\
\{ G_a , G_b \} & = & D_{ab}\,^c G_c \, ,
\end{eqnarray}
with  $\{ f , g\} = \Omega^{\mu\nu} \frac{\partial f}{\partial
x^{\mu}}\frac{\partial g}{\partial x^{\nu}}$ and $\{ x^{\mu} , x^{\nu} \}=
\Omega^{\mu\nu}$ provided that the gauge transformation of any function $F:
\Gamma_{ext} \rightarrow R$ is the same for both cases
\begin{eqnarray}
\delta_{\varepsilon} F = \varepsilon^a \{ F , \gamma_a \}_0 = \varepsilon^a \{
F , G_a \} \, . \label{QQPP}
\end{eqnarray}
Note that, by construction, the {\it evolution} in $\tau$ of $x^{\mu}$ is {\it
not} modified by the choice of (\ref{XXMM}) or (\ref{CCC}) simply because
$\tau$-evolution is a gauge transformation, i.e., the explicit form of this
transformation is the same for both cases. In fact, from (\ref{XXMM})
\begin{eqnarray}
x^{\mu} (\tau + d\tau) & = & x^{\mu}(\tau) + {\dot x}^{\mu} d \tau = x^{\mu}
(\tau) + \varepsilon^a \{ x^{\mu} , \gamma_a \}_0 = x^{\mu}(\tau) +
\varepsilon^a \{ x^{\mu} , G_a \} \, ,
\end{eqnarray}
with $\varepsilon^a = \lambda^a dt$ and last equality follows from Eq.
(\ref{QQPP}).

In some particular cases, the $\gamma$'s in (\ref{XXMM}) will form a Lie
algebra. However, it might be possible that the $G$'s in (\ref{CCC}) form a
Lie algebra, {\it distinct} to that of the $\gamma$'s or even that the $G$'s
do {\it not} form a Lie algebra. In addition, suppose that there exists a set
of observables (same for both cases) which form a Lie algebra with respect to
the Poisson brackets of Eq. (\ref{XXMM}). They might or might not form a Lie
algebra with respect to the Poisson brackets of Eq. (\ref{CCC}). Moreover, it
might or not be possible to find a set of observables that form a Lie algebra
with respect to the Poisson brackets of Eq. (\ref{CCC}).

\section{Concluding remarks}
The covariant canonical formalism applied to parametrized nonrelativistic
Hamiltonian systems clearly displays the various geometrical structures
involved in their dynamics. In particular, the reduced phase space is reached
by using Dirac's observables which are the constants of motion with respect to
$t$ of the standard (i.e., non-parametrized) Hamiltonian system. In contrast
to what happens in Dirac's method, in the covariant canonical formalism there
is not need to choose a gauge condition to get the reduced phase space. In
spite of using the techniques of the covariant canonical formalism to analyze
the geometry of parametrized nonrelativistic Hamiltonian systems, the usual
symplectic structure was used.

To avoid the use of the usual symplectic structure in the extended phase space
$\Gamma_{ext}$, it was explored what changes in Dirac's canonical formalism if
alternative symplectic structures are chosen. It was shown that there exists
the freedom to choose the symplectic structure on the extended phase space if
the equation that defines the constraint surface is, in the generic case,
accordingly modified in such a way that the gauge transformation is not
altered. Moreover, due to the fact that the null vectors are the same for any
choice of the pair $(\Omega_{ext}, G)$ where $\Omega_{ext}$ is the symplectic
structure on $\Gamma_{ext}$ and $G\approx 0$ defines the constraint surface
$\Sigma$ then the reduced phase space $\Gamma_{phys}$ is also not modified,
what changes is the symplectic structure $\Omega_{phys}$ on $\Gamma_{phys}$
which depends on the pair $(\Omega_{ext}, G)$ chosen. The generalization of
these results to any generally covariant systems with a finite number of first
class constraints was also discussed.

Finally, due to the fact that the canonical analysis is a first step towards
canonical quantization and because it was seen how the algebra of observables
directly depends on the symplectic structure chosen in the extended phase
space then it would interesting to know how this phenomenon (i.e., the fact of
choosing distinct symplectic structures and distinct ways of expressing the
constraints surface) is handled, for instance, in Dirac's quantization as well
as in the framework of algebraic and/or refined algebraic quantizations which
heavily depend on the algebra of observables.

\section*{Acknowledgements}
Warm thanks to G.F. Torres del Castillo and J. D. Vergara for their comments
and criticisms to the first version of this paper. M. Montesinos also
acknowledges support from the {\it Sistema Nacional de Investigadores} (SNI)
of the {\it Consejo Nacional de Ciencia y Tecnolog\'{\i}a} (CONACyT) of
Mexico.


\end{document}